\documentclass[sn-aps,iicol]{sn-jnl}

\usepackage{graphicx}%
\usepackage{multirow}%
\usepackage{amsmath,amssymb,amsfonts}%
\usepackage{amsthm}%
\usepackage{mathrsfs}%
\usepackage[title]{appendix}%
\usepackage{xcolor}%
\usepackage{textcomp}%
\usepackage{manyfoot}%
\usepackage{booktabs}%
\usepackage{algorithm}%
\usepackage{algorithmicx}%
\usepackage{algpseudocode}%
\usepackage{listings}%

\usepackage{upgreek}

\theoremstyle{thmstyleone}%
\theoremstyle{thmstyletwo}%

\theoremstyle{thmstylethree}%

\begin{document}

\title{Short-Range Tests of the Gravitational Inverse-Square Law}

\author*[1]{ \fnm{Jiro} \sur{Murata} }\email{jiro@rikkyo.ac.jp}
\author[1,2]{ \fnm{Takuhiro} \sur{Fujiie} }
\author[1]{ \fnm{Sae} \sur{Suzuki} }
\affil[1]{\orgdiv{Department of Physics}, \orgname{Rikkyo University}, \orgaddress{\city{Tokyo}, \postcode{171-8501}, \country{Japan}}}
\affil[2]{\orgdiv{Present address: Department of Physics}, \orgname{Konan University}, \orgaddress{\city{Kobe}, \postcode{658-8501}, \country{Japan}}}

\abstract{Experimental constraints on the gravitational inverse-square law at short range are presented, employing a consistent formalism across a wide range of length scales. We provide comprehensive updates incorporating new experimental data from the past decade---spanning tabletop forces, Casimir measurements, slow neutron scattering, and collider searches---building upon our previous review. This work facilitates the direct comparison of experimental results with theoretical models that extend general relativity. Furthermore, a comparison between various model parametrizations, including extra-dimensional models, is introduced. Finally, results from tabletop experiments are compared with those from high-energy collider experiments for both Yukawa and power-law potentials.}





\maketitle


\section{Introduction}\label{introduction}
Testing gravity is testing spacetime.
Newton's gravitational inverse-square law stands as the founding monument of modern science; it remains valid for most gravitational phenomena in our universe, except in extreme conditions where modifications from general relativity are necessary.
The inverse-square law has been tested with great precision, to be sure, but such high-precision tests have been performed only on planetary scales.
On the other hand, precision in short-range tests at sub-laboratory scales remains poor.
To address this, the primary objective of this paper is to provide a comprehensive, updated review of short-range gravity experiments, building upon our previous review from a decade ago \cite{Murata_2015}. This update incorporates the remarkable experimental progress achieved over this past decade across multiple frontiers, including high-precision tabletop force measurements, slow neutron scattering, atomic spectroscopy, and high-energy collider experiments.

Since the end of the last century, searching for deviations from the inverse-square law at microscales has attracted significant interest, especially in light of the possible hypothesis of extra spatial dimensions beyond our four-dimensional spacetime.
Before then, several claims, such as the fifth-force hypothesis, also prompted numerous experimental searches for violations of the inverse-square law from the 1970s to today.
A wide variety of experimental reports can be found in such attempts, from the geophysical scale down to the quark scale.
It is not always straightforward to compare their results with theoretical models, as each experiment reports its results using different parametrizations.
In this paper, we aim to summarize the wide variety of experimental results within the same parameter spaces and using the same criteria, while distinguishing directly published constraints in the literature from our reconstructed or reinterpreted limits.

Specifically, we examine the current bounds and sensitivities of the Yukawa-type coupling strength over each relevant length scale---ranging from high-precision tabletop torsion balances that constrain the dimensionless coupling strength in the micrometer-to-millimeter regime, to Casimir force measurements at sub-micrometer scales, and slow neutron scattering and high-energy collider experiments probing the sub-atomic regime. This systematic comparison is motivated by the hierarchy problem in extra-dimensional scenarios \cite{Arkani.Hamed1998263, PhysRevLett.83.3370, PhysRevLett.83.4690}. These theoretical frameworks have driven outstanding progress not only in high-energy collider experiments \cite{PhysRevD.103.112006, CMS2024}  but also in recent tabletop experiments, spanning high-precision torsion balances \cite{PhysRevLett.124.101101, PhysRevLett.124.051301}, Casimir force measurements \cite{PhysRevLett.116.221102}, and slow neutron scattering \cite{doi:10.1126/science.abc2794}.

The remainder of this paper is organized as follows. Section~\ref{observables} establishes the model-independent parameters used to characterize potential deviations from Newtonian gravity. 
Section~\ref{ed} provides the theoretical framework for extra-dimensional scenarios, discussing power-law, Kaluza--Klein, and Yukawa potentials.
Section~\ref{Yukawa-par} outlines the general properties of the Yukawa parametrization and the mechanics of absolute versus relative measurements. Section~\ref{constraints} presents the unified experimental constraints and explicitly details the dual translation workflows used to map various experimental results into consistent parameter spaces representing Yukawa-type forces, power-law modifications, and higher-dimensional Planck scales. Section~\ref{discussion} provides a comparative discussion of these boundaries, Section~\ref{experiments} reviews the key updated experiments, and Section~\ref{conclusion} concludes the paper.


\section{Observables and Parametrization}\label{observables}
If an experiment measures a gravitational potential $V(r)$ or a force $F(r)$ between two point masses $m_1$ and $m_2$ at a separation distance $r$, these quantities are compared with the Newtonian gravitational laws, $V_{\rm N}(r)$ or $F_{\rm N}(r)$:
\begin{eqnarray}
V_{\rm N}(r) &=& -G_{\rm N} \frac{m_1 m_2}{r}, \nonumber \\*
F_{\rm N}(r) &=& G_{\rm N} \frac{m_1 m_2}{r^2},
\label{Newton}
\end{eqnarray}
where $G_{\rm N}$ is Newton's gravitational constant. To characterize potential deviations, we introduce the following parameters in our work \cite{Murata_2015}:
\begin{eqnarray}
\Gamma &\equiv& \frac{V(r)}{V_{\rm N}(r)}, \quad \Delta \equiv \Gamma - 1, \nonumber \\
\gamma &\equiv& \frac{F(r)}{F_{\rm N}(r)}, \quad \delta \equiv \gamma - 1.
\label{gamma-delta}
\end{eqnarray}
Observing $\Delta \neq 0$ or $\delta \neq 0$ indicates that the Newtonian gravitational law is violated at the experimental distance $r$.
In the following, we basically use $\Delta$ for theoretical formulations and $\delta$ for experimental analyses.
Although $\Delta$ and $\delta$ are closely related, they differ slightly due to the spatial derivative:
$
\delta = \Delta - r \frac{d\Delta}{dr}
$.
Therefore, this second term must be included when analyzing experimental data from force measurements.

Experiments test whether $\delta = 0$, while theoretical models predict non-zero values based on their specific parameters.
In practice, actual experiments cannot employ ideal point masses; therefore, the net force must be estimated via volume integrations.
Because a single value of $\delta(r)$ can represent only point-mass configurations, most experimental works report their boundaries using theoretical model parameters rather than the model-independent parameter $\delta(r)$.
This parametrization difference is precisely why direct comparison and physical interpretation can become confusing when different models are used.

Note that there is no theoretically known value of $G_{\rm N}$.
Its determination was first made possible by Cavendish in 1798 by directly measuring $F(r)$ with known $m_1, m_2$, and $r \sim r_{\rm N} \equiv 0.1\text{ m}$ values, using a torsion balance bar \cite{10.1098/rstl.1798.0022}.
To this day, $G_{\rm N} = 6.708\,83(15) \times 10^{-39} \, \hbar c \,({\rm GeV}/c^2)^{-2}$ is measured in similar Cavendish-type experiments, improving its relative precision to $\sim 10^{-5}$ \cite{ParticleDataGroup:2024cfk}.

\begin{figure*}[t]
 \begin{center}
  \includegraphics[width=100mm]{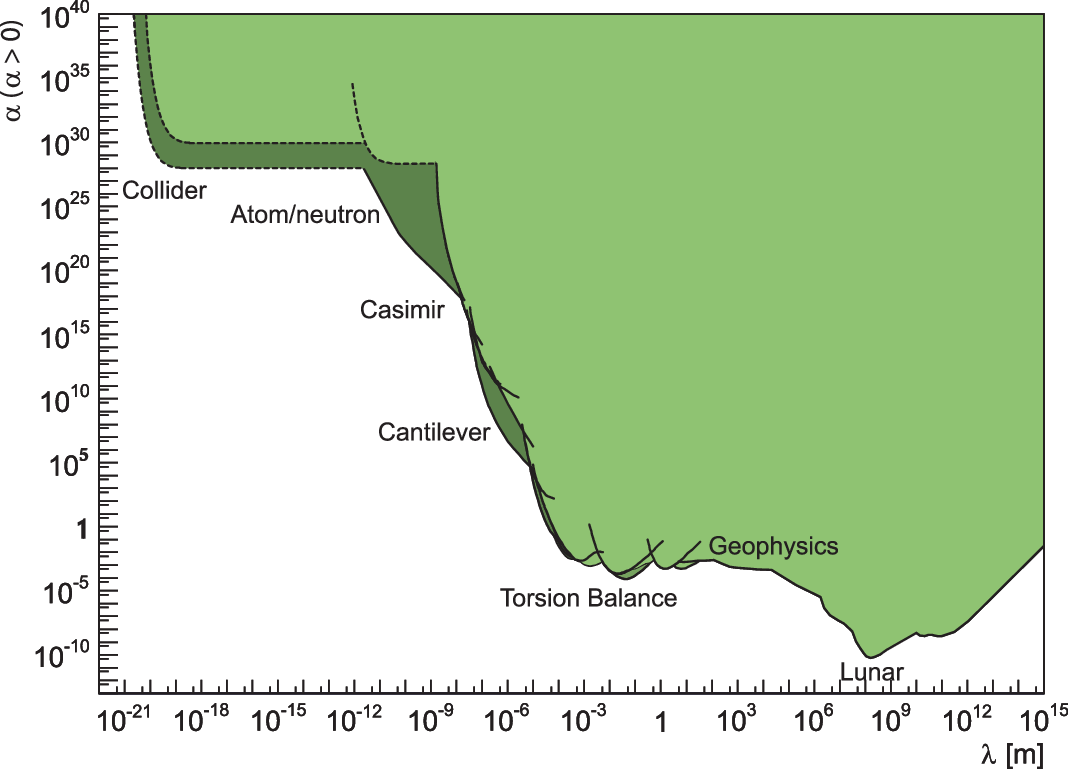}
 \end{center}
\caption{$\alpha$--$\lambda$ plot in the Yukawa parametrization as defined in Eq.~(\ref{V_al}). Experimental constraints are shown as shaded areas at the 95\% confidence level, with dark-shaded regions representing the improved constraints obtained over the last decade, following Ref.~\cite{Murata_2015}. 
Solid lines denote direct experimental results, while dashed lines indicate estimates from our work.}
 \label{alpha-lambda-fullscale}
\end{figure*}

There are two representative parametrizations for testing the inverse-square law experimentally. The first one is a power-law parametrization expressed as
\begin{eqnarray}
V_{\rm power}(r)&=&V_{\rm N}(r)
\left[
	1+\left( \frac{\Lambda}{r} \right) ^n 
\right] \nonumber \\ 
(n, \Lambda&:&\rm {model \;\; parameters}),
\label{V_nl}
\end{eqnarray}
the other one is a Yukawa parametrization,
\begin{eqnarray}
V_{\rm Yukawa}(r)&=&V_{\rm N}(r)
\left[
	1+\alpha \; {\rm exp} \left( - \frac{r}{\lambda} \right)  
\right] \nonumber \\ 
(\alpha, \lambda&:&\rm {model\;\; parameters}).
\label{V_al}
\end{eqnarray}
These models suppose possible new short-range potentials, which are effective only at $r \lesssim \Lambda$ or $\lambda$, 
in addition to the Newtonian gravitational potential $V_{\rm N}$.
Here, $\Lambda$ and $\lambda$ are the scale parameters where the modification from $V_{\rm N}$ is going to be significant.
$n$ and $\alpha$ represent the power and strength of the additional potentials, respectively.
For example, the constraint curves in the $\alpha$--$\lambda$ parameter space are shown from the planetary scale to the quark scale in Fig.~\ref{alpha-lambda-fullscale}.
This $\alpha$--$\lambda$ plot is the most commonly used expression for the experimental results, showing $\alpha$'s excluded region for supposing values of $\lambda$.
In this paper, we restrict our scope to attractive interactions with positive $\alpha$, which directly correspond to extra-dimensional scenarios. While constraints on repulsive interactions ($\alpha < 0$) can be derived similarly and exhibit qualitatively analogous experimental sensitivities, they are omitted in our figures to maintain visual clarity.

The inverse-square law is based on the assumption that our world is an isotropic three-dimensional space, as implied by Gauss's law.
On the other hand, if our space contains isotropic $n$-extra dimensions, the potential shape must be modified as $1/r \rightarrow 1/r^{1+n}$.
Equation~(\ref{V_nl}) represents such extra-dimension models.

The Yukawa potential (\ref{V_al}) represents more conservative models, which predict new boson-exchange forces with the reduced Compton wavelength $\lambda=\frac{\hbar}{m_{\rm b} c}$, where $m_{\rm b}$ is the mass of the exchanging boson.
This new interaction needs to couple to the masses $m_1$ and $m_2$ in the same way as gravity does.
Or, other quantities $q\sim m$, such as baryon number $B$, can be the corresponding ``charge'' for this additional Yukawa interaction.
If we do not distinguish between this charge $q$ and mass $m$, gravitational phenomena that seem to violate the universality of free fall can be expected \cite{Ninomiya_2017}.

As we see here, testing the inverse-square law is closely related to the search for extra dimensions, new bosons, and violations of the equivalence principle.
High-energy collider experiments align with the power-law parametrization, whereas other short-range laboratory experiments usually prefer using the Yukawa 
parametrization. 
However, it should be quite simple to compare their results with each other if all of them can be expressed in $\delta (r)$ as model-independent parameters.
In this work, we relate the power-law parameters $n$--$\Lambda$ and the Yukawa parameters $\alpha$--$\lambda$ to each other based on this idea, following the approach in Ref.~\cite{Murata_2015}.
We provide updated constraints on these parameter spaces by incorporating recent short-range experimental data. 
It should be noted that astrophysical constraints are beyond the scope of this study.
For detailed reviews and comprehensive discussions on these indirect astrophysical and stellar  constraints, we refer the reader to the classic work~\cite{Raffelt1996} and recent literature~\cite{Fiorillo2025}.

\section{Interpretation in the Context of Extra Dimensions}\label{ed}
Most experimental attempts over the past quarter of a century have been motivated by theoretical predictions of deviations from the gravitational inverse-square law in brane-world scenarios.
Therefore, it is instructive to discuss these parametrizations in the context of extra-dimensional models.
Among these, the large extra-dimension model (the Arkani-Hamed--Dimopoulos--Dvali, or ADD, model) \cite{Arkani.Hamed1998263} and the warped extra-dimension model (the Randall--Sundrum, or RS, model) \cite{PhysRevLett.83.3370, PhysRevLett.83.4690} represent two notable predictions.
Indeed, both the ADD and RS-II models predict power-law-type deviations from Newtonian gravity at short distances.
Historically, the concept of a compactified extra dimension was proposed in Kaluza--Klein (KK) theory a century ago \cite{Klein1926}. 
Since then, the compactification scale had traditionally been assumed to be the Planck length,
\begin{equation}
L_{\rm Pl}=\sqrt{ \frac {\hbar G_{\rm N}}{c^3}  },
\end{equation}
making it far too small to be probed experimentally.

However, it is important to note that the gravitational constant used to estimate the Planck length is the experimental value $G_{\rm N}$ measured at laboratory scales $r_{\rm N}$, which is conventionally assumed to remain constant down to $L_{\rm Pl}$.
The central concept of the ADD model is that the effective gravitational coupling varies significantly with the length scale.
In this framework, the fundamental higher-dimensional Planck scale is dramatically reduced from the four-dimensional Planck mass $M_{\rm Pl}$ down to the TeV scale, where gravity and the Standard Model interactions can share a comparable energy scale, thereby naturally resolving the hierarchy problem.

It is useful to interpret the power-law parametrization in Eq.~(\ref{V_nl}) and the Yukawa parametrization in Eq.~(\ref{V_al}) in the context of extra-dimensional theories.
Here, we discuss these formulations specifically within the framework of the ADD model, which posits the existence of $n$ additional compactified spatial dimensions whose characteristic scale is significantly larger than the Planck length $L_{\rm Pl}$.
If all the extra dimensions are isotropic and share a common characteristic scale, the gravitational potential obeys the $1/r^{1+n}$ power law at distances shorter than this scale, whereas the standard Newtonian inverse-square law holds at larger distances, as qualitatively represented by the power-law parametrization.
Specifically, the modified gravitational potential $V_{\rm ADD}(r)$ is expressed in terms of the higher-dimensional gravitational constant $G_{4+n}$ and the phenomenological transition scale $\Lambda$ in the short-range limit ($r \ll \Lambda$) as
\begin{eqnarray}
V_{\rm ADD}(r) &=& -G_{4+n} \frac{m_1 m_2}{r^{1+n}} \nonumber \\*
&=& -G_{\rm N} \frac{m_1 m_2}{r} \left( \frac{\Lambda}{r} \right)^n ,
\end{eqnarray}
which yields
\begin{eqnarray}
G_{4+n} &=& G_{\rm N} \Lambda^n  \nonumber \\
\Rightarrow \;\;\; M_{\rm Pl}^2 &=& M_{\rm D}^{2+n} \Lambda^n (c/\hbar)^n ,
\label{ADD}
\end{eqnarray}
where $M_{\rm Pl}$ and $M_{\rm D}$ are the Planck masses corresponding to $G_{\rm N}$ and $G_{4+n}$, respectively. 
This algebraic relation can be determined by dimensional analysis; thus, Eq.~(\ref{ADD}) serves as the fundamental definition of the higher-dimensional Planck mass $M_{\rm D}$ for a given phenomenological transition scale $\Lambda$, independent of specific extra-dimensional geometries.

The strong power-law dependence of $1/r^{1+n}$ leads to a much stronger gravitational interaction at $r \ll \Lambda$ compared to the standard Newtonian potential $V_{\rm N}$.
The original motivation for the ADD model is that the fundamental higher-dimensional Planck scale $M_{\rm D}$ can be significantly lower than the four-dimensional Planck mass $M_{\rm Pl}$, potentially reaching the TeV scale and thereby naturally resolving the hierarchy problem.
Using the relation in Eq.~(\ref{ADD}), the power-law parametrization conventionally expressed in the $n$--$\Lambda$ plane can be mapped directly onto the $n$--$M_{\rm D}$ parameter space.
Consequently, constraints from high-energy collider experiments expressed in $n$--$M_{\rm D}$ and results from tabletop force measurements expressed in $n$--$\Lambda$ can be directly compared within a unified framework.

It should be noted that the transition scale parameter $\Lambda$ in Eq.~(\ref{V_nl}) is defined phenomenologically as the scale where the potential deviates from Newtonian gravity; thus, it does not a priori require a specific geometric interpretation of the extra-dimensional space. 
However, when physically interpreting this transition parameter $\Lambda$ as the geometric size of the compactified extra dimensions, two different conventions coexist in the literature: the compactification radius definition ($\Lambda \equiv R$) \cite{Arkani.Hamed1998263} and the torus circumference definition ($\Lambda \equiv 2\pi R$) commonly adopted in high-energy collider publications to represent the volume of the extra dimensions \cite{ParticleDataGroup:2024cfk}. 

To avoid model-dependent ambiguities, we consistently plot all constraints in the subsequent figures using the phenomenological transition scale $\Lambda$ directly, as defined in Eqs.~(\ref{V_nl}) and (\ref{ADD}). 

On the other hand, to facilitate convenient and direct comparisons with different conventions in the existing literature, we explicitly discuss the physical implications under both the radius and circumference interpretations when presenting specific numerical limits in the text.

The next point to consider is the relationship between the power-law and the Yukawa parametrizations, which represent fundamentally different physical models.
\par
Nevertheless, laboratory experiments conventionally report their results in the $\alpha$--$\Lambda$ parameter space, even when motivated by searches for extra dimensions.
To resolve this apparent contradiction, the conceptual framework of extra-dimensional models can be utilized as follows.
Below, we present a simple illustrative discussion as a representative example.

The energy-momentum relation for a massless particle in a spacetime with extra dimensions is given by
\begin{eqnarray}
E^2 &=& \left[ (p_1 c)^2 + (p_2 c)^2 + (p_3 c)^2 \right] \nonumber \\
&& \quad + \left[ (p_4 c)^2 + (p_5 c)^2 + \cdots \right] .  
\end{eqnarray}
If the particle has momentum components along the extra dimensions, such as the ``missing momentum'' invisible from our four-dimensional spacetime, it is observed as an effective non-zero mass $M$ satisfying
$(Mc^2)^2 = (p_4 c)^2 + (p_5 c)^2 + \cdots$.
Here, $p_i$ denotes the momentum component along the $i$-th spatial dimension.
The corresponding de Broglie wavelength along the fourth dimension is $l_4 = h/p_4$, for example.
Considering a compactified extra dimension with radius $R$, a periodic boundary condition can be applied along this dimension.
Analogous to Bohr's atomic model, the de Broglie wavelength along the extra-dimensional circumference $2\pi R$ must be quantized as $l_4 = 2\pi R/m$, where $m = 0, 1, 2, 3, \dots$ is the quantum number.
This boundary condition directly translates to the quantization of the momentum component as $p_4 = mh/(2\pi R) = m\hbar/R$.
Assuming this particle is a graviton, a quantum of graviton mass is naturally defined as $M_0 = \frac{\hbar}{R c}$.
In this compactified extra-dimensional model, the graviton mass $M_m$ can be expressed as an integer multiple of this unit mass, $M_m = m M_0$.
These modes, $M_m = 0, M_0, 2M_0, 3M_0, \dots$, are referred to as the Kaluza--Klein (KK) tower.
Each graviton with mass $M_m$ behaves as a massive boson mediating a gravitational-like force.
Their interaction ranges are naturally expressed as Yukawa potentials with their reduced Compton wavelength $\lambda_m \equiv \frac{\hbar}{M_m c} = \lambda/m$, where we identify the Yukawa scale parameter $\lambda$ with the compactification radius $R$ ($\lambda = R$).
This mechanism is of particular interest regarding the dynamical origin of mass \cite{HOSOTANI1983309}.

By employing the Yukawa scale parameter $\lambda$ defined above, the total gravitational potential $V_{\text{KK}}(r)$, representing the sum over all KK modes, can be phenomenologically expressed in terms of the individual coupling strengths $\alpha_m$ of each massive mode as
\begin{equation}
V_{\text{KK}}(r)/V_{\text{N}}(r) = 1 + \sum_{m=1}^{\infty} \alpha_m e^{-mr/\lambda},
\end{equation}
where the parameter $\lambda$ consistently corresponds to the compactification radius $R$.
\par
In specific extra-dimensional models, these coupling coefficients $\alpha_m$ are not arbitrary free parameters but can be calculated based on the geometric configurations and dimensions of the compactified space. For example, in a simple isotropic torus-compactified scheme with $n$ extra dimensions, the leading Yukawa coefficient is predicted to be $\alpha_1 = 2n$, which accounts for both the spatial degeneracy of the extra dimensions and the two-directional degrees of freedom (clockwise and counter-clockwise propagation) along each compactified dimension \cite{Kehagias200039}.

This multi-mode summation yields two distinct physical regimes depending on the scale of the measurement:
{\renewcommand{\labelitemi}{--} 
\begin{itemize}
    \item Long-range regime ($r \gg \lambda$): All massive KK modes ($m \geq 1$) are exponentially suppressed ($e^{-mr/\lambda} \to 0$). Consequently, only the massless graviton mode ($m=0$) survives, and the standard four-dimensional Newtonian potential $V_{\text{N}}(r)$ is naturally recovered. In this regime, the single Yukawa parametrization (Eq.~(\ref{V_al})) serves as a good first-order approximation by keeping only the lightest massive mode ($m=1$) and neglecting the heavily suppressed higher-mass modes ($m \geq 2$).
    \item Short-range regime ($r \ll \lambda$): The exponential suppression of the higher-mass KK modes is lifted, and a vast number of these modes contribute coherently to the force. This collective summation of Yukawa potentials effectively smooths out into the higher-dimensional power-law potential, behaving as $V_{\text{KK}}(r) \propto r^{-(1+n)}$.
\end{itemize}
}

This dual behavior highlights a critical limitation of the single Yukawa parametrization (Eq.~(\ref{V_al})) when testing extra-dimensional theories. Because a single Yukawa term with a constant $\alpha$ cannot capture the rapid, multi-mode algebraic growth that occurs at $r \ll \lambda$, interpreting experimental data at a distance scale $r$ far below the target interaction range $\lambda$ using a simple Yukawa approximation will artificially distort the reconstructed coupling strength.

Therefore, to compare tabletop experiments and high-energy collider results in a consistent and model-independent manner across vastly different distance scales, we recommend directly utilizing the $n$--$\Lambda$ or $n$--$M_{\rm D}$ power-law parameter spaces, which naturally encompass the multi-mode characteristics of extra-dimensional models.


\section{Properties of Yukawa Parametrization}\label{Yukawa-par}
\begin{figure*}[t]
 \begin{center}
  \includegraphics[width=100mm]{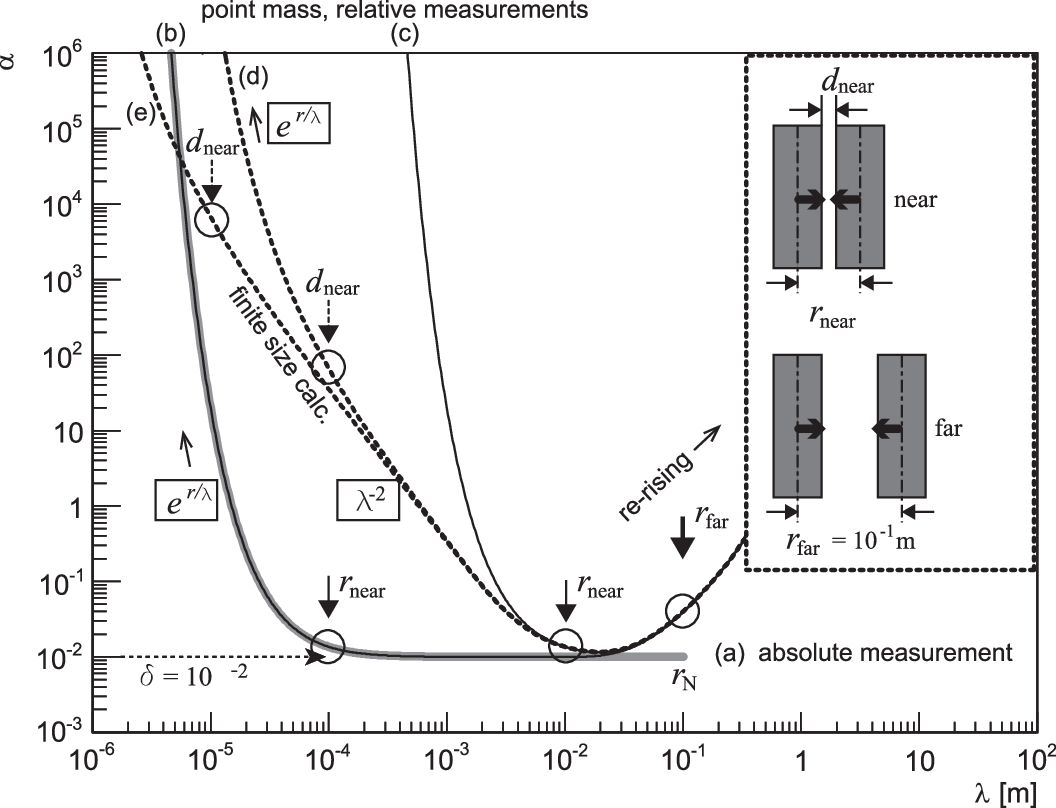}
 \end{center}
 \caption{Typical examples of the $\alpha$--$\lambda$ plot in the Yukawa parametrization for model cases. (a) Absolute measurement: point-mass calculation, where $r_{\rm N}$ can be regarded as $r_{\rm far}$ in relative measurements. The bottom line corresponds to the experimental precision on $\delta$. The exponential rise $e^{r/\lambda}$ toward small $\lambda$ is significant at $\lambda < r_{\rm near}$. (b) and (c): Results of relative measurements for point masses at $r_{\rm near} = 10^{-2}$~m and $10^{-4}$~m with $r_{\rm far} = 10^{-1}$~m. In addition to the exponential rise at small $\lambda$, a re-rising feature at $\lambda > r_{\rm far}$ toward large $\lambda$ is confirmed in relative measurements. (d) and (e): Results for finite-sized, parallel-plate masses, where the gap distance $d_{\rm near}$ and the center-to-center distance $r_{\rm near}, r_{\rm far}$ are defined as in the inset figure. They exhibit the exponential rise $e^{r/\lambda}$ at $\lambda < d_{\rm near}$ toward small $\lambda$. The intermediate region at $d_{\rm near} < \lambda < r_{\rm near}$ shows straight-line features, exhibiting an $\alpha \propto \lambda^{-2}$ dependence.}
 \label{al-model}
\end{figure*}

Before discussing the experimental results, the general properties of the Yukawa parametrization should be outlined.
If we use the Yukawa parametrization in Eq.~(\ref{V_al}) to analyze the experimental data $\Delta(r)$, $\alpha$ can be experimentally obtained for a given setting parameter $\lambda$ using
\begin{equation}
\Delta=\frac{V(r)}{V_{\rm N}(r)}-1=\alpha \; e^{-r/\lambda},
\end{equation}
which leads to
\begin{equation}
\alpha=\Delta \; e^{r/\lambda} .
\label{al_exp}
\end{equation}
If the result on $\Delta$ is consistent with zero, the upper limit of $\Delta$ primarily determines the upper limit on $\alpha$.
Experimental uncertainty of $r$ also contributes to that of $\alpha$, but it is usually relatively small.
In the case of force measurements, 
\begin{eqnarray}
\delta&=&\frac{F(r)}{F_{\rm N}(r)}-1=\alpha \left( 1+ \frac{r}{\lambda}\right) e^{-r/\lambda}   \nonumber \\
\Rightarrow \quad
\alpha&=&\frac{\delta}{1+r/\lambda} e^{r/\lambda} \nonumber \\ 
&\sim& \delta \, e^{r/\lambda} \quad (r\ll \lambda),
\label{al_exp_abs}
\end{eqnarray}
which gives approximately the same functional shape of $\alpha (\lambda)$ as Eq.~(\ref{al_exp}).
The functional shape of the constraint curve on the $\alpha$--$\lambda$ plot should be dominated by the simple exponential shape of $e^{r/\lambda}$ with its bottom at $\alpha = \delta$.
Fig.~\ref{al-model}(a) shows an example of the functional shape.

The above discussion assumes the simplest case of point-mass calculation using the known value of $G_{\rm N}$, which we call ``absolute measurements'' \cite{Murata_2015}.
In this case, the uncertainty of $G_{\rm N}$ or $F_{\rm N}$ also contributes to the total uncertainty of $\delta$ as
\begin{equation}
\left( \frac{\sigma_\delta}{\delta} \right)^2=
\left( \frac{\sigma_F}{F} \right)^2+\left( \frac{\sigma_{F_{\rm N}}}{F_{\rm N}} \right)^2 ,
\end{equation}
where, here and in the following, $\sigma_X$ denotes the uncertainty of a quantity $X$ in general.
Then, if the experimental relative precision of $F$, $\frac{\sigma_F}{F}$, is expected to be smaller than that of $F_{\rm N}$, $\frac{\sigma_{F_{\rm N}}}{F_{\rm N}}$, one should avoid employing this absolute measurement based on the literature value of $G_{\rm N}$.
Otherwise, the total precision will be dominated by the uncertainty of $G_{\rm N}$.

Instead, ``relative measurements'' do not rely on the absolute value of $G_{\rm N}$ to avoid its relatively poor precision.
They perform measurements at a minimum of two different distances $r = r_{\rm near}$ and $r_{\rm far}$ to cancel $G_{\rm N}$. 
As described in detail in Ref.~\cite{Murata_2015}, the relative measurements combine the results $\delta(r_{\rm near})$ and $\delta(r_{\rm far})$ to cancel $G_{\rm N}$; the resulting functional shapes are shown in Figs.~\ref{al-model}(b) and (c).

A feature of these relative measurements is the re-rising of $\alpha$ at large $\lambda$ ($\lambda > r_{\rm far}$).
The constraint curve has a minimum point, unlike the simple exponential curve in Eq.~(\ref{al_exp_abs}) for absolute measurements.
Most short-range laboratory experiments employ this technique and typically report asymmetric, convex constraint curves.
Now, the simple exponential curve for absolute measurements in Eq.~(\ref{al_exp_abs}) can be understood as the case of $r_{\rm far} = r_{\rm N}$ relative measurements using $G_{\rm N}$ as the far-position data.

It is also useful to discuss the finite-size effects of the test and source masses. In such cases, the total potential must be integrated over the volume elements of each object. As detailed in the calculations of Ref.~\cite{Murata_2015}, there are two characteristic length scales: the distance $r$ between the centers of mass and the gap distance $d$ between the nearest surfaces. Examples of these calculations for parallel-plate geometries are shown in Fig.~\ref{al-model}.
In the region where $\lambda < d$, we observe a simple exponential increase at small $\lambda$, similar to the behavior in absolute measurements (\ref{al_exp_abs}). 
However, a key feature emerges in the intermediate region, $d < \lambda < r$: the constraint curve is no longer exponential but follows a power-law behavior, appearing as a straight line on a logarithmic scale with $\alpha \propto \lambda^{-2}$, as seen in Figs.~\ref{al-model}(d) and (e).

A measurement involving finite-sized objects can be treated as a superposition of contributions from microscopic volume elements at various separation distances. While the volume elements near the gap surface provide an exponential constraint for $\lambda < d$, elements deeper within the objects contribute to the force at relatively larger distances. This extends the effectiveness of the constraint into the $d < \lambda$ region, up to the limit where $\lambda \sim r_{\rm near}$. Consequently, the characteristic behavior $\alpha \propto \lambda^{-2}$ can be interpreted as a ``dilution effect'' arising from the spatial extension of the objects.

Regardless of the type of measurement, the small-$\lambda$ region shows the same exponential characteristics as the absolute measurement.
This is straightforward to understand because, at a distance below the minimum experimental distance, there is no experimental data $\delta(r)$.
Consequently, sensitivity to test the Yukawa potential is rapidly lost because only its exponential tail can be measured.

In this section, we saw that there should be the following key features expected in the $\alpha$--$\lambda$ parameter plots:
{\renewcommand{\labelitemi}{--} 
\begin{itemize}
\item Feature~1: $\alpha$'s bottom line represents the experimental precision of $\delta$.
\item Feature~2: Simple exponential rising towards smaller $\lambda$ at $\lambda<r$.
\item Feature~3: Convex shape around the measurement distances $r$ for relative measurements, which shows a re-rising behavior towards large $\lambda$ at $\lambda>r_{\rm far}$.
\item Feature~4: $\alpha\propto\lambda^{-2}$ constraint arising from the finite sizes of the objects.
\end{itemize}
}

\section{Experimental Constraints}\label{constraints}
To facilitate a unified and direct comparison, the constraint curves and data points plotted in our figures are categorized into three distinct types, which are visually distinguished by different line styles: 
(i) original limits taken directly from the respective experimental publications, 
(ii) our own phenomenological estimations reconstructed from published experimental uncertainties where explicit parameterized curves were unavailable, and 
(iii) translated limits converted from high-energy collider results using the theoretical relations of the ADD model \eqref{ADD}.

While we plot these constraints on the same axes to facilitate a unified comparison, we caution the reader that the directly published limits and our reinterpreted estimates have different statistical origins. The former represent the confidence boundaries at a 95\% confidence level reported by the experimental collaborations, whereas the latter are illustrative reconstructions based on reported experimental uncertainties and point-mass approximations.

In the following, we present the $\alpha$--$\lambda$ constraints using the published results from the original literature wherever they are available.
When direct results are unavailable, we evaluate the characteristic parameters $\delta(r)$ for typical experiments and plot them as our own estimates based on Eq.~\eqref{al_exp_abs}.
Since most experiments do not report their results in terms of power-law parameters, we derive the corresponding power-law constraints by determining a representative $\delta(r)$ for each experiment that reproduces the published $\alpha$--$\lambda$ exclusion curve, assuming the relation in Eq.~\eqref{al_exp_abs}.
Using these extracted $\delta(r)$ parameters, the $n$--$\Lambda$ and $n$--$M_{\rm D}$ plots are constructed as phenomenological constraints for the tabletop experiments, as described in Section~\ref{Power-pot}.
We clarify that these conversions are based on a simplified point-mass framework using the extracted effective $\delta(r)$ parameters. While the original experimental exclusion curves inherently account for finite-size effects in their analyses, these detailed physical corrections are not explicitly modeled or individually recalculated in our algebraic translations.

For the high-energy collider constraints, we directly adopt the published lower limits on $M_{\rm D}$ to place the data points in the $n$--$M_{\rm D}$ plot.
These limits are then converted into the $n$--$\Lambda$ plot using the theoretical relation in Eq.~\eqref{ADD}.
Finally, the corresponding collider limits in the $\alpha$--$\lambda$ plane are reconstructed by employing the $\delta(r)$ values that best reproduce these $n$--$\Lambda$ results.

\subsection{Yukawa Potential}\label{Yukawa-pot}
\begin{figure*}[t]
 \begin{center}
  \includegraphics[width=100mm]{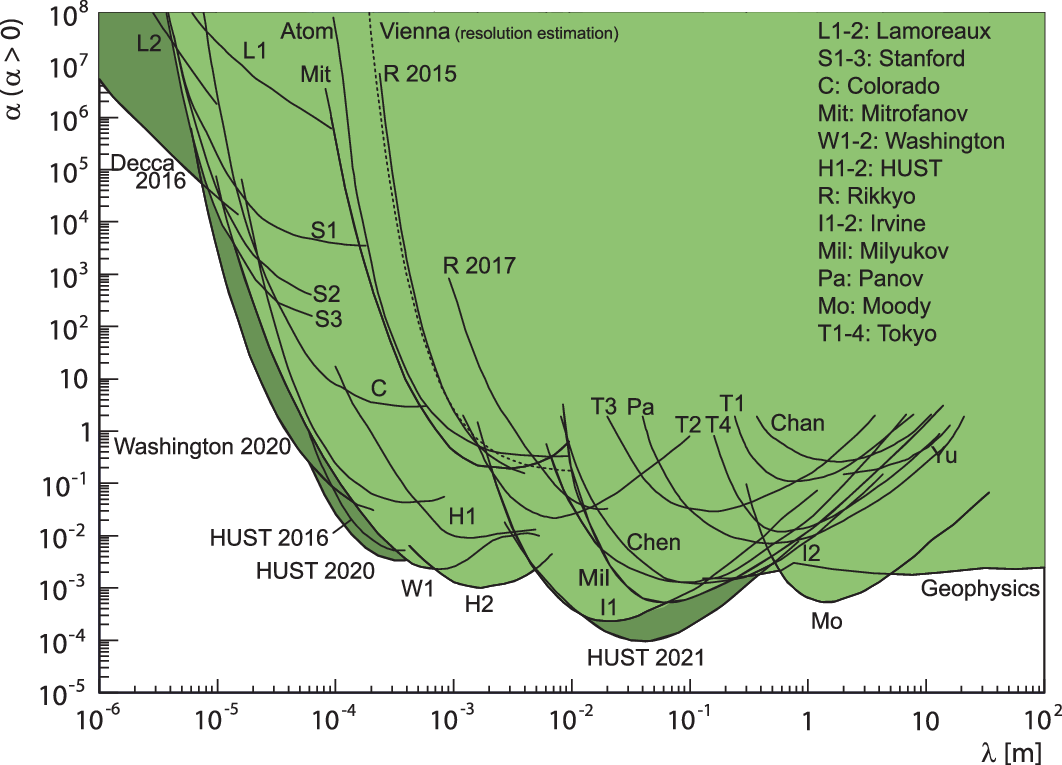}
 \end{center}
 \caption{$\alpha$--$\lambda$ plot in the Yukawa parametrization for geophysics at $100$~m to short-range laboratory experiments at micrometers. 
The dark shaded area shows the improved constraints obtained over the last decade, following Ref.~\cite{Murata_2015}. 
New constraints are given by the UW (Washington 2020 \cite{PhysRevLett.124.101101}) and the HUST groups (HUST 2016 \cite{PhysRevLett.116.131101}, HUST 2020 \cite{PhysRevLett.124.051301}, HUST 2021 \cite{PhysRevLett.126.211101}). New measurements are reported using conventional torsion balance bars from the Vienna \cite{Vienna2021} and the Rikkyo groups (R 2015 \cite{Murata_2015}, R 2017 \cite{Ninomiya_2017}) and using atomic interferometry (Atom \cite{Panda2024}). 
Solid and dotted lines represent direct experimental results and our phenomenological estimations reconstructed using the experimental spatial resolutions, respectively.
All other results shown in this figure are summarized in Ref.~\cite{Murata_2015} and the references therein.
}
 \label{alpha-lambda-shortscale}
\end{figure*}

We are now ready to examine the experimental results. Figure~\ref{alpha-lambda-shortscale} presents the $\alpha$--$\lambda$ plot for the short-range region, spanning from $10^{-6}$~m to $100$~m. 
All experiments shown here were performed using mechanical measurement techniques.
The best precision of $\alpha < 10^{-4}$ was obtained at centimeter scales \cite{PhysRevLett.126.211101}, implying a measurement precision of $\delta < 10^{-4}$ (Feature~1). 
Most experiments conducted at scales larger than one millimeter exhibit the characteristic re-rising of $\alpha$ typical of relative measurements at large $\lambda$ (Feature~3). In contrast, smaller-scale experiments report simple exponential constraints (Feature~2). 
This behavior arises because their precision $\delta$ is lower than the known precision of $G_{\rm N}$; in such cases, the uncertainty in $G_{\rm N}$ is relatively small compared to that of the measured force $F$.

Most results obtained at scales below 1~cm were reported following the proposal of the ADD model, with the primary objective of searching for $\alpha \neq 0$. 
In particular, numerous experiments have been conducted at the micrometer-to-millimeter scale, motivated by the ADD prediction of $\lambda \sim 0.1$~mm for $n = 2$ and $M_{\rm D} \sim 1$~TeV. 
This region has remained of significant interest for exploring large extra-dimension models.
Figure~\ref{alpha-lambda-shortscale} also includes recent updates from the decade since Ref.~\cite{Murata_2015}. 
Sophisticated torsion balance experiments by the University of Washington (UW) and the Huazhong University of Science and Technology (HUST) groups have spearheaded exploration in this regime.
Both groups have continued to report steady progress throughout this decade. 
Detailed descriptions of their experimental setups are provided in Section~\ref{experiments}.

\begin{figure*}[t]
 \begin{center}
  \includegraphics[width=100mm]{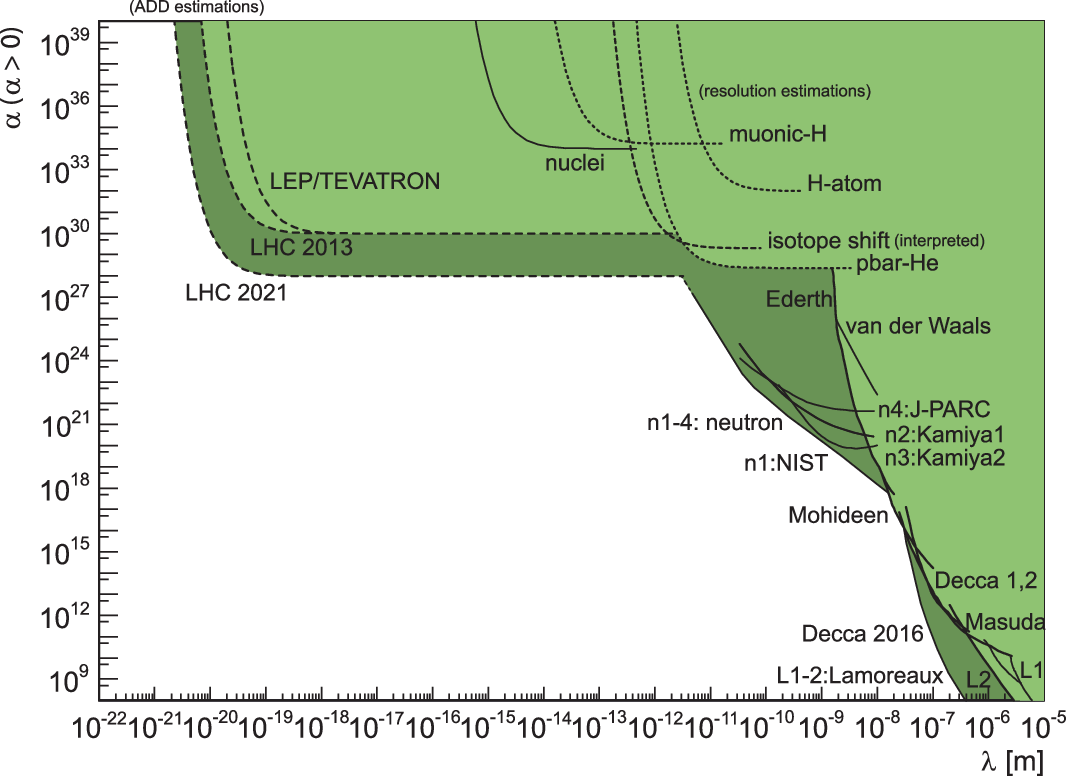}
 \end{center}
 \caption{$\alpha$--$\lambda$ plot in the Yukawa parametrization for sub-micrometer scales. 
The dark shaded area shows the improved constraints obtained over the last decade, following Ref.~\cite{Murata_2015}. 
New constraints are provided by a Casimir force experiment (Decca~2016 \cite{PhysRevLett.116.221102}), neutron scattering experiments (NIST \cite{doi:10.1126/science.abc2794}, J-PARC \cite{PhysRevD.97.062002}, and Kamiya~1 \cite{PhysRevLett.114.161101}--2 \cite{10.1063/5.0036985}), and LHC experiments (LHC~2013 \cite{PhysRevLett.110.011802}--2021 \cite{PhysRevD.103.112006}). 
Additional new constraints are reported from atomic spectroscopy (isotope shifts \cite{PhysRevLett.134.233002, PhysRevLett.134.063002, Kyoto-Yb}). 
Solid, dotted, and dashed lines denote direct experimental results, phenomenological estimations reconstructed from the experimental spatial resolutions, and estimations converted from high-energy collider results based on the ADD model, respectively.
All other results shown in this figure are summarized in Ref.~\cite{Murata_2015} and the references therein.}
 \label{alpha-lambda-microscale}
\end{figure*}

Considering that the KK theory predicts $\alpha \sim \mathcal{O}(1)$ or $\alpha = 2n$ with $n = 2$ as a particular case \cite{Kehagias200039}, we can roughly exclude this model at $\lambda > 10^{-5}$~m based on these results from the UW data.
However, since the prediction $\alpha = 2n$ is theoretically valid only in the regime where the KK potential is well-represented by a single, leading Yukawa term (i.e., $\lambda \sim r$), we do not apply this extra-dimensional interpretation across the entire $\alpha$--$\lambda$ plot, where higher KK modes dominate the short-range limit ($r \ll \lambda$).
As discussed, interpreting the results of the extra-dimension model in this $\alpha$--$\lambda$ plot can be misleading; therefore, we will examine them in more detail using the power-law parametrization in the next section.

Figure~\ref{alpha-lambda-microscale} shows the $\alpha$--$\lambda$ plot in the micrometer range, down to the quark scale.
Many of the experiments contributing to this plot were not intended to test gravity.
Some results in the region $\lambda > 10$~nm are derived by interpreting the corresponding results from Casimir force experiments using mechanical apparatuses.

In contrast to short-range laboratory experiments, implementing an electromagnetic shield between the test and source masses becomes increasingly challenging. Consequently, experimental uncertainties are often dominated by electromagnetic couplings. Neutron experiments, which have seen significant improvements in sensitivity over the past decade, aim to suppress these couplings by employing a neutral probe. At scales below those accessible by neutron scattering, gravity tests must rely on systems consisting of charged particles.

In atomic systems, where electromagnetic forces dominate, any potential gravitational contribution must be estimated within the bounds of theoretical uncertainty. High-energy collider experiments attempt to detect gravity-related phenomena more directly; however, these events are also dominated by the vast Standard Model background. Thus, the precision of Standard Model predictions dictates the overall sensitivity, a situation analogous to that found in atomic systems.

Since many previous studies did not report their results in terms of the $\alpha$--$\lambda$ parametrization, several curves in these figures are derived in the present work and are indicated by dotted lines in Figs.~\ref{alpha-lambda-shortscale}--\ref{alpha-lambda-microscale}. 
In particular, interpreting the high-energy collider data requires a specific analysis within the power-law framework, and the resulting converted constraints are indicated by dashed lines.

As shown in these figures, significant improvements have been achieved in Casimir force measurements, neutron scattering, and high-energy collider experiments over the past decade. 
These will be further discussed within the context of the power-law parametrization in the next section.
Figure~\ref{alpha-lambda-fullscale} additionally presents constraints at planetary scales. 
\par
While observations of gravitational waves from binary neutron stars and black holes can, in principle, provide quantitative constraints on deviations from the inverse-square law, current published analyses do not yet offer competitive constraints at shorter scales \cite{PhysRevD.105.064061, PhysRevD.97.064039}.

In summary, the most stringent constraints in the $\alpha$--$\lambda$ plane across the vast range of length scales are dominated by the following specific experimental frontiers:
{\renewcommand{\labelitemi}{--} 
\begin{itemize}
    \item $\lambda < 10\text{ pm}$ (sub-atomic scale): LHC collider experiments (e.g., ATLAS monojet results \cite{PhysRevD.103.112006});
    \item $10\text{ pm} < \lambda < 10\text{ nm}$ (nanometer scale): slow neutron scattering experiments, spearheaded by the NIST group \cite{doi:10.1126/science.abc2794};
    \item $10\text{ nm} < \lambda < 1\ \mu\text{m}$ (sub-micrometer scale): Casimir force measurements, with the strongest limits set by Decca et al. \cite{PhysRevLett.116.221102};
    \item $1\ \mu\text{m} < \lambda < 1\text{ mm}$ (micrometer-to-millimeter scale): high-precision tabletop torsion balance experiments by the UW \cite{PhysRevLett.124.101101} and HUST \cite{PhysRevLett.124.051301} groups;
    \item $1\text{ mm} < \lambda < 1\text{ m}$ (millimeter-to-meter scale): classical torsion balance measurements by the new HUST result \cite{PhysRevLett.126.211101};
    \item $\lambda > 1\text{ m}$ (large scale): geophysical and astronomical observations \cite{Murata_2015}.
\end{itemize}
}

\subsection{Power-Law Potential}\label{Power-pot}

\begin{figure*}[t]
 \begin{center}
  \includegraphics[width=100mm]{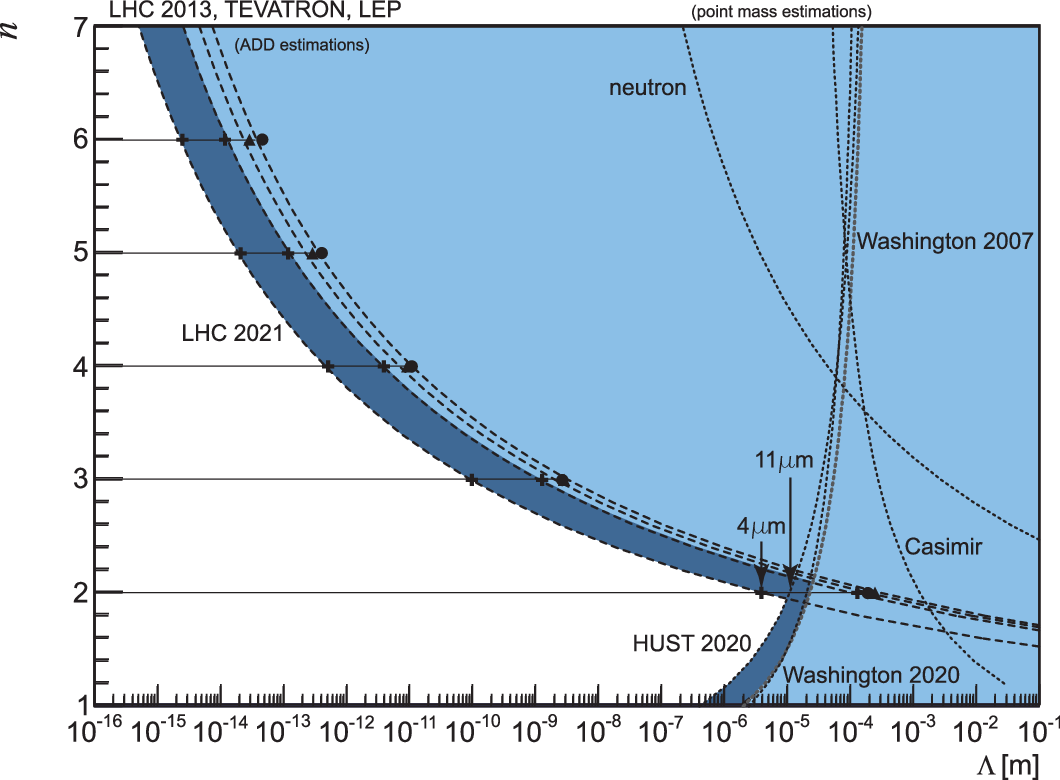}
 \end{center}
\caption{$n$--$\Lambda$ plot in the power-law parametrization as defined in Eq.~(\ref{V_nl}). We estimated the values of $\delta(r)$ that reproduce the reported results in the $\alpha$--$\lambda$ plots of the representative laboratory experiments to obtain these $n$--$\Lambda$ constraints.
The dark shaded area shows the improved constraints obtained over the last decade, following Ref.~\cite{Murata_2015}. 
Results for the UW group (Washington 2007 \cite{PhysRevLett.98.021101}--2020 \cite{PhysRevLett.124.101101}) and the HUST group (HUST 2020 \cite{PhysRevLett.124.051301}) give the tightest constraints for $n=2$ among all short-range laboratory experiments (Casimir: Decca 2016 \cite{PhysRevLett.116.221102}, neutron: NIST \cite{doi:10.1126/science.abc2794}). 
Results for high-energy collider experiments (LHC 2021 \cite{PhysRevD.103.112006} and \cite{Murata_2015}) are obtained by interpreting the reported $M_{\rm D}$ into $\Lambda$ using the ADD model Eq.~(\ref{lambda-MD}). 
The lines represent the power-law functional forms in Eq.~(\ref{n-l-absolute}). 
Dotted and dashed lines denote phenomenological estimations reconstructed from point-mass calculations with estimated $\delta(r)$ parameters for typical tabletop experiments, and limits converted from high-energy collider results based on the ADD model, respectively.
Parameters $\delta(r)$ for the collider experiments (LHC 2013--2021, TEVATRON, LEP) are estimated in this plot to reproduce the $\Lambda(n)$ data points using Eq.~(\ref{n-l-absolute}).
All data points and limits are plotted using the phenomenological transition scale parameter $\Lambda$ of Eq.~(\ref{V_nl}) without model-dependent geometric assumptions. Under this consistent parameter space, the HUST limit is at $11\,\upmu\text{m}$ and the LHC limit is at $4\,\upmu\text{m}$. If these are physically interpreted under the torus compactification scheme where $\Lambda$ represents the circumference ($2\pi R$), the corresponding physical compactification radius limits are $R \approx 1.75\,\upmu\text{m}$ and $0.64\,\upmu\text{m}$, respectively.
}
 \label{n-lambda}
\end{figure*}

The power-law parametrization can directly represent the extra-dimension models.
However, most of the experimental results have been reported in the Yukawa parametrization.
In this study, we estimated the effective $\delta(r)$ for the reported experiments to approximate their results in the power-law parametrization.
For simplicity, we use the point-like framework to estimate the values of $\delta(r_{\rm near})$ and $\delta(r_{\rm far})$ that approximately reproduce the reported $\alpha$--$\lambda$ curves for short-range laboratory experiments \cite{Murata_2015}. 
Similar to the Yukawa coupling range $\alpha(\lambda)$ in Eq.~\eqref{al_exp_abs}, the relation $\Lambda(n)$ can be estimated for the power-law potential using $\delta$ for absolute measurements at $r$ as
\begin{equation}
\delta=(1+n) \left( \frac{\Lambda}{r} \right)^n .
\label{delta_absolute}
\end{equation}
For relative measurements, the relationship between $n$ and $\Lambda$ can be obtained as
\begin{eqnarray}
\Lambda&=&\left(
\frac{ \delta/(1+n)
}{
1-(\delta+1)(r_{\rm near}/r_{\rm far})^n
}
\right)^{1/n} r_{\rm near} \label{n-l-relative} \\
\rightarrow&&
\left(\frac{\delta}{1+n}\right)^{1/n} r \quad  (r \equiv r_{\rm near} \ll r_{\rm far}). \label{n-l-absolute}
\end{eqnarray}
Equation~\eqref{n-l-relative} represents the relationship for relative measurements as discussed in Ref.~\cite{Murata_2015}, and Eq.~\eqref{n-l-absolute} corresponds to the limit for absolute measurements, consistent with Eq.~\eqref{delta_absolute}. 
We use this simplified Eq.~\eqref{n-l-absolute} to illustrate the constraint curves.

For short-range laboratory experiments, we estimated the characteristic parameters $\delta(r)$ from the $\alpha$--$\lambda$ curves of each experiment, which set the tightest constraints on $\alpha$ at the corresponding $\lambda$.
The results are shown in Fig.~\ref{n-lambda} for several typical experiments, which are updated from the same figure in Ref.~\cite{Murata_2015}.
In Fig.~\ref{n-lambda}, results from high-energy collider experiments are also shown as $\Lambda(n)$ data points, obtained by interpreting the reported $M_{\rm D}$ values.
We can use Eq.~\eqref{ADD} to convert $M_{\rm D}$ data into $\Lambda$ values as
\begin{equation}
\Lambda=\frac{(M_{\rm Pl} c^2)^{2/n}}{(M_{\rm D} c^2)^{1+(2/n)}}\hbar c.
\label{lambda-MD}
\end{equation}
Here, we use Einstein's reduced gravitational constant $\kappa = 8\pi G_{\rm N}/c^4$ to numerically estimate the reduced Planck mass $M_{\rm Pl} = \sqrt{\hbar/\kappa c^3} = 2.43 \times 10^{15}$~TeV in this work.
The collider data points in the $n$--$\Lambda$ plot shown in Fig.~\ref{n-lambda} were directly estimated using Eq.~\eqref{lambda-MD}.
We also estimated the corresponding $\delta(r)$ for the collider results to reproduce the experimental $\Lambda(n)$ data points using the functional form in Eq.~\eqref{n-l-absolute}, shown as the dashed lines in Fig.~\ref{n-lambda}.
We can see good agreement between the data points and the model functions of Eq.~\eqref{n-l-absolute}.
These estimated $\delta(r)$ parameters are then used to reconstruct the $\alpha$--$\lambda$ curves for the collider results.

\begin{figure*}[t]
 \begin{center}
  \includegraphics[width=100mm]{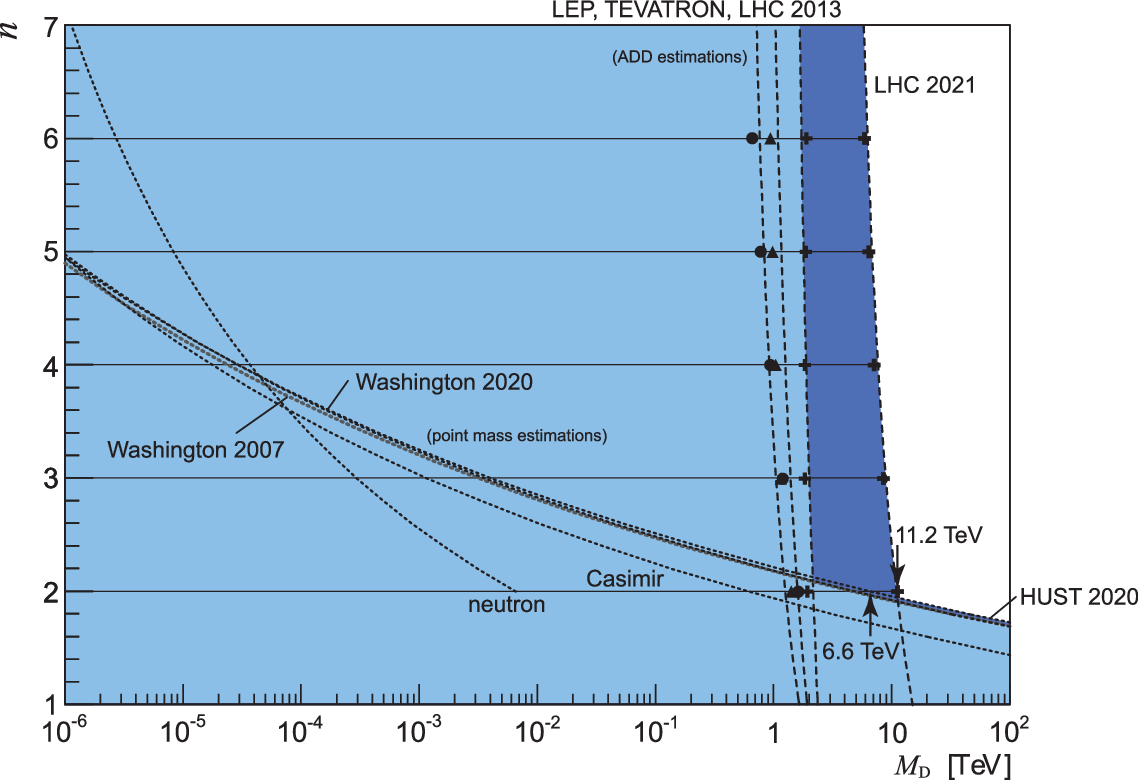}
 \end{center}
 \caption{
$n$--$M_{\rm D}$ plot expressing the ADD model parameters corresponding to Fig.~\ref{n-lambda}.
The dark shaded area shows the improved constraints obtained over the last decade, following Ref.~\cite{Murata_2015}. 
$M_{\rm D}$ data points for the collider experiments are their reported values.
Dotted lines denote phenomenological estimations of $M_{\rm D}$ reconstructed via Eq.~\eqref{MD-lambda} using the functional form of Eq.~\eqref{n-l-absolute} for representative laboratory experiments.
Dashed lines denote the $n$--$M_{\rm D}$ curves estimated for the collider experiments.
The limits are consistently plotted using the phenomenological relation of Eq.~\eqref{n-l-absolute} directly from the transition scale parameter $\Lambda$ shown in Fig.~\ref{n-lambda}, independent of specific model-dependent geometric assumptions. 
Under this unified definition, HUST and the LHC set competitive limits of $M_{\rm D} > 6.6$~TeV and $M_{\rm D} > 11.2$~TeV for $n = 2$, respectively. 
}
 \label{n-MD}
\end{figure*}

Conversely, $\Lambda$ can be converted into $M_{\rm D}$ using Eq.~\eqref{ADD} as
\begin{equation}
M_{\rm D} c^2 =\left[
(M_{\rm Pl} c^2)^{2/n} \frac{\hbar c}{\Lambda}
\right]^\frac{1}{1+2/n}.
\label{MD-lambda}
\end{equation}
The $n$--$\Lambda$ curves for the short-range laboratory experiments shown in Fig.~\ref{n-lambda} are also translated into $M_{\rm D}$ limits using Eq.~\eqref{MD-lambda}, which are represented as dotted lines in the $n$--$M_{\rm D}$ plot in Fig.~\ref{n-MD}.
The $n$--$\Lambda$ curves for the high-energy collider data are similarly converted into the $n$--$M_{\rm D}$ limits, shown as dashed lines in Fig.~\ref{n-MD} along with the officially reported $M_{\rm D}$ data points \cite{Murata_2015}.
Here, we again observe good agreement between the collider data points $M_{\rm D}(n)$ and the model functions.
This agreement demonstrates that the rough estimation using Eq.~\eqref{n-l-absolute} well represents the official ADD analysis results.

The $n$--$\Lambda$ (Fig.~\ref{n-lambda}) and $n$--$M_{\rm D}$ (Fig.~\ref{n-MD}) plots provide direct representations of the constraints on the large extra-dimension model \cite{Arkani.Hamed1998263}. 
These figures demonstrate that recent LHC and HUST experiments have significantly advanced the upper limits at the forefront of the search for extra dimensions. 
For the case of $n = 2$, the upper limit on $\Lambda$ has been tightened to $4\,\upmu\text{m}$, while the lower limit on $M_{\rm D}$ has been pushed up to $11.2$~TeV \cite{PhysRevD.103.112006}. 
This progress is notable when compared to the previous benchmarks of $\Lambda < 23\,\upmu\text{m}$ and $M_{\rm D} > 4.6$~TeV \cite{Murata_2015}.


\section{Discussion}\label{discussion}
The $\alpha$--$\lambda$ plots (Figs.~\ref{alpha-lambda-shortscale} and \ref{alpha-lambda-microscale}), along with the $n$--$\Lambda$ (Fig.~\ref{n-lambda}) and $n$--$M_{\rm D}$ (Fig.~\ref{n-MD}) plots, clearly demonstrate the significant progress achieved over the past decade. 
Both micrometer-scale precision experiments and high-energy collider studies provide stringent constraints on the large extra-dimension hypothesis. 
In particular, the increased statistics and upgraded energies at the LHC have substantially improved the limits on the Yukawa coupling strength $\alpha$ at sub-atomic scales.

Figures~\ref{n-lambda} and \ref{n-MD} illustrate that for the specific case of $n = 2$, micrometer-scale laboratory experiments can effectively compete with high-energy searches. 
For $n > 2$, however, high-energy collider experiments remain the only effective approach to exploring deviations from the gravitational inverse-square law.
Beyond these established methods, numerous experimental advancements have emerged over this past decade. Many of these have successfully imposed new constraints on the $\alpha$--$\lambda$ plane, further excluding the exchange of hypothetical new bosons in these regimes. 
Several novel experimental efforts \cite{Vienna2021, Panda2024, Ninomiya_2017, PhysRevLett.134.233002, PhysRevLett.134.063002, Kyoto-Yb} are also underway; although they do not yet establish the most stringent limits, they represent promising technologies for future exploration. 


Regarding gravitational wave observations, while the finite size of stellar objects may eventually allow for constraints at sub-planetary scales, current reports remain relatively weak \cite{PhysRevD.105.064061, PhysRevD.97.064039}. 
In general, gravitational wave observations are unlikely to achieve the precision required to test short-range gravity at or below the laboratory scale.


\section{List of Experiments}\label{experiments}
Recent experiments, whose results are shown in the constraint plots, are introduced in this section.

\subsection{Torsion Balance Experiments}

Since Cavendish's era, the torsion balance has remained the most promising instrument for measuring extremely weak forces in short-range gravity experiments. 
As illustrated in Figs.~\ref{n-lambda} and \ref{n-MD}, results from the UW and HUST groups are uniquely competitive with the LHC constraints in the $n=2$ case.

\begin{figure}[H]
 \begin{center}
  \includegraphics[width=75mm]{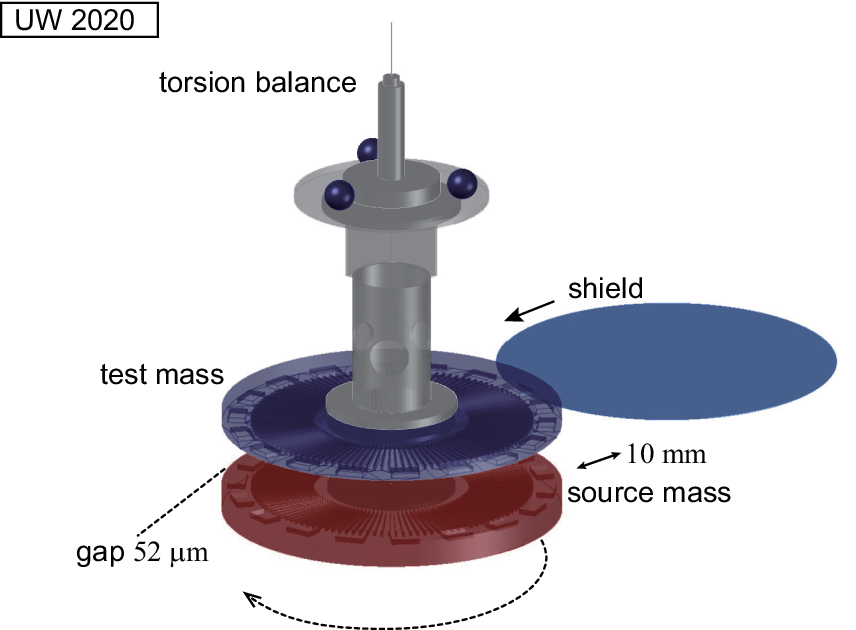}
 \end{center}
 \caption{Experiment of the UW group \cite{PhysRevLett.124.101101}, following their previous works \cite{PhysRevLett.98.021101, PhysRevLett.86.1418, PhysRevD.70.042004}. 
Test mass (18- and 120-fold, $54\,\upmu\text{m}$ thick, $5.5$~cm diameter platinum disks) suspended by an $83$~cm long wire; source masses (18- and 120-fold, $99\,\upmu\text{m}$ thick, $5.5$~cm diameter platinum disks) are placed at a $d_{\rm near} = 52\,\upmu\text{m}$--$3.0$~mm (surface-to-surface) gap. 
An electrostatic shield ($10\,\upmu\text{m}$ gold-coated beryllium-copper membrane) is set between the test and source masses.
The source mass rotates around the vertical axis.
}
 \label{UW2020}
\end{figure}

\begin{figure}[H]
 \begin{center}
  \includegraphics[width=75mm]{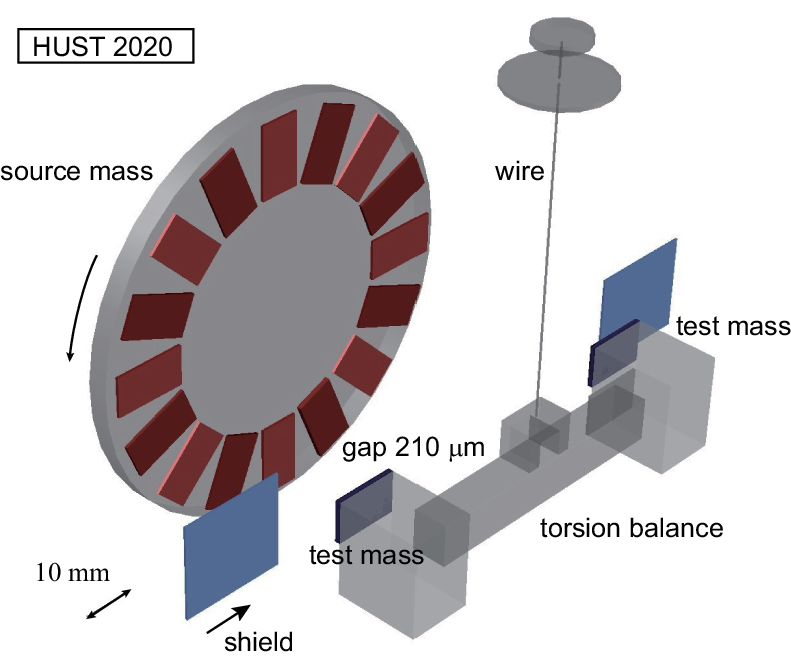}
 \end{center}
 \caption{Experiment of the HUST group \cite{PhysRevLett.124.051301}, following their previous works \cite{PhysRevLett.116.131101, PhysRevLett.108.081101, PhysRevLett.98.201101}. 
Test mass (two $0.2$~mm thick, $14.6\,\text{mm} \times 12.0\,\text{mm}$ wide tungsten plates) attached to a torsion pendulum 
using a wire (tungsten, $25\,\upmu\text{m}$ diameter, $70$~cm long).
Source near mass: $1.8$~mm thick, $20.8\,\text{mm} \times 20.8\,\text{mm}$ wide tungsten plate 
at a $d_{\rm near} = 210\,\upmu\text{m}$ (surface-to-surface) gap. 
The source mass (eight $0.2$~mm thick, $17.6\,\text{mm} \times 11.4\,\text{mm}$ wide tungsten plates attached to a $100$~mm diameter disk) rotates around the horizontal axis. 
The torsion balance position is stabilized using feedback from capacitive actuators. 
An electrostatic shield ($30\,\upmu\text{m}$ beryllium-copper membrane) is set between the test and source masses.
 }
 \label{HUST2020}
\end{figure}

The UW group has led this field for decades, with their latest results \cite{PhysRevLett.124.101101} obtained using the apparatus shown in Fig.~\ref{UW2020}. 
They achieved measurements with a minimum gap distance of $d = 52\,\upmu\text{m}$. 
Similarly, the HUST group has consistently tested the inverse-square law using centimeter-scale torsion balances. 
Their most recent result \cite{PhysRevLett.124.051301}, shown in Fig.~\ref{HUST2020}, utilized a gap distance of $d = 210\,\upmu\text{m}$ --- an improvement from their previous setup at $d = 295\,\upmu\text{m}$ \cite{PhysRevLett.116.131101}. 
Both groups employ electrostatic shields between the test and source masses to mitigate systematic couplings.

The University of Vienna \cite{Vienna2021} and Rikkyo University \cite{Ninomiya_2017} groups have also reported millimeter-scale results using similar torsion balance configurations equipped with electrostatic shields.
While these did not establish the most stringent constraints at their target scales (Fig.~\ref{alpha-lambda-shortscale}), the 2017 Rikkyo experiment was specifically designed to test the composition dependence of $G$, providing the most stringent constraints on the universality of free fall at the shortest ranges.
Furthermore, a novel experiment by the University of California, Berkeley group utilized atom interferometry \cite{Panda2024} at this scale; although its current sensitivity does not yet surpass existing limits (Fig.~\ref{alpha-lambda-shortscale}), it represents a promising new approach.

While UW, HUST, Rikkyo, and Berkeley provided their results in the $\alpha$--$\lambda$ parametrization, many other studies did not. 
Consequently, the dotted curves in Figs.~\ref{alpha-lambda-shortscale} and \ref{alpha-lambda-microscale} represent our own estimations derived from published experimental data. 
For the Vienna group, we converted their reported precision of $\delta \sim 0.1$ \cite{Vienna2021} into the corresponding constraint.
Notably, at the $10\,\upmu\text{m}$ scale, the Stanford group's result \cite{PhysRevD.78.022002} was the sole experimental effort until the latest UW result. 
Below the micrometer scale, implementing electrostatic shielding becomes increasingly prohibitive, while the Casimir force begins to emerge as a dominant and challenging background.

\subsection{Casimir Force Experiments}

\begin{figure}[H]
 \begin{center}
  \includegraphics[width=75mm]{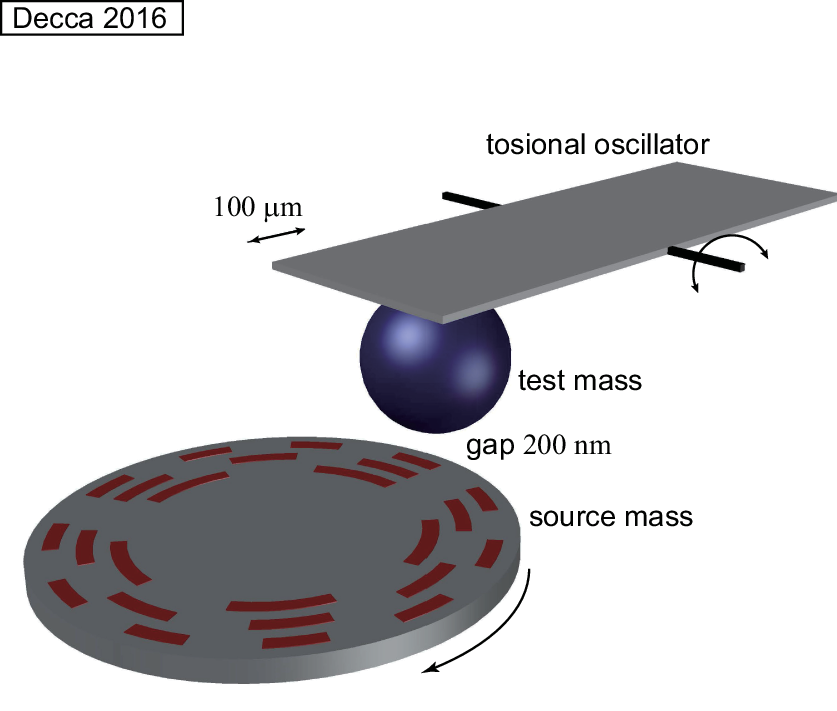}
 \end{center}
 \caption{
Experiment of Decca et al. (Indiana-Purdue group) \cite{PhysRevLett.116.221102} following their previous works \cite{PhysRevLett.94.240401,PhysRevD.75.077101}. 
Test mass: $149.3\,\upmu\text{m}$ radius sapphire sphere (chromium and gold coated) attached to a microelectromechanical torsional oscillator ($500\,\upmu\text{m} \times 500\,\upmu\text{m}$).
Source masses ($2.1\,\upmu\text{m}$ thick, $150\,\upmu\text{m}$ wide gold or silicon layer with a common $150$~nm thick gold layer coating) on a disk rotate around the axis.
The gap was set at $d = 200$--$1000$~nm (surface-to-surface). 
No shielding is applied between the test and source masses.
This experiment is designed to cancel the Casimir force ambiguity by measuring the force difference between silicon and gold to search for strong gravity.
}
 \label{Decca2016}
\end{figure}

At the sub-micrometer scale, Casimir force measurements are primarily conducted to verify QED calculations. By re-analyzing these results, we can estimate the maximum allowable strength of gravitational couplings, even though the sensitivity is often limited by the theoretical uncertainties of the Casimir force itself. 
The Decca group at Indiana-Purdue University \cite{PhysRevLett.116.221102} has been specifically testing gravitational couplings using dedicated devices designed to minimize the Casimir contributions. 
Their latest result, obtained with the setup shown in Fig.~\ref{Decca2016}, provides the most stringent constraint at around the $100$~nm scale (Fig.~\ref{alpha-lambda-microscale}), representing a three-order-of-magnitude improvement over their previous study \cite{PhysRevLett.94.240401}.

The $\alpha$--$\lambda$ exclusion curves for Casimir-type experiments exhibit a characteristic power-law behavior, $\alpha \propto \lambda^{-2}$. 
This feature, previously identified as Feature~4 in Fig.~\ref{al-model}, can be attributed to the finite-size effects of the test and source masses. Indeed, while the minimum gap distance in the setup of Fig.~\ref{Decca2016} is $d = 200$~nm, the radius of the test mass is $149.3\,\upmu\text{m}$, which is much larger than $d$. 
Thus, finite-size effects are substantial in this configuration.

Mechanical fabrication at the micrometer scale and below is increasingly challenging with current technologies, making Casimir-type measurements the practical limit for mechanical gravity tests. 
Below the $10$~nm scale, experimental efforts must shift to atomic, molecular, and sub-atomic systems, despite the dominance of electromagnetic interactions in these regimes.

\subsection{Atomic and Sub-Atomic Experiments}

Atomic systems serve as ``micro-laboratories'' for testing interactions between nuclei and orbital electrons. 
Due to the extraordinary precision with which atomic energy levels can be measured, coupled with the ability to perform rigorous QED calculations of the electromagnetic interactions, these systems provide a unique platform for gravity tests.
The sensitivity to gravitational contributions is determined by the interplay between experimental precision and theoretical accuracy. 
However, despite this remarkable precision, the immense ratio of the Coulomb force to the gravitational force limits the sensitivity to approximately $\alpha < 10^{30}$.

As demonstrated in torsion-balance experiments, suppressing electromagnetic forces significantly enhances sensitivity to $\alpha$. 
While the Casimir force cannot be shielded in experiments, the gravitational signal can still be extracted by designing sophisticated apparatuses that effectively cancel the Casimir contributions. 
Without such mechanical means to suppress electromagnetic couplings, the sensitivity for gravity tests inevitably degrades exponentially, as previously discussed. 
Figure~\ref{alpha-lambda-microscale} presents estimated constraints for several atomic systems. It should be noted that most of these are not direct experimental reports but are instead our own estimations derived from combining experimental and theoretical uncertainties (represented by $\delta$), including those of the relevant physical constants \cite{Murata_2015}.

At the nanometer scale, a series of neutron scattering experiments has established constraints on the $\alpha$--$\lambda$ plane \cite{doi:10.1126/science.abc2794, PhysRevD.97.062002, PhysRevLett.114.161101, 10.1063/5.0036985}; these results are summarized in Fig.~\ref{alpha-lambda-microscale}. 
In contrast to atomic spectroscopy, these studies measure the neutron-scattering cross sections of atomic targets. Utilizing the neutron as a probe allows for significant suppression of electric interactions, which would otherwise dominate the signals in atomic systems.
Within a wave-mechanical framework, these experiments probe potential mass-coupling interactions, such as non-Newtonian gravity. In this context, the impact parameter $b$ corresponds to the characteristic distance $r$ used in mechanical measurements, while the spatial resolution is fundamentally limited by the de Broglie wavelength of the incident neutron beam. 
The coupling strength $\alpha$ is thus derived from the observed scattering cross sections.
Notably, the constraint reported by the NIST group provides the most stringent limit in this regime \cite{doi:10.1126/science.abc2794}. In Figure.~\ref{alpha-lambda-microscale}, instead of utilizing the classical real-space exponential growth, we extrapolate their constraint (indicated by the dotted line) by applying a $\lambda^{-4}$ algebraic power-law dependence as described in Ref.~\cite{doi:10.1126/science.abc2794}.

This power-law scaling is grounded in quantum scattering theory: neutron scattering probes the Fourier transform of the potential, where a Yukawa interaction yields a momentum-dependent amplitude proportional to $\lambda^2 / (1 + q^2\lambda^2)$. In the short-range limit ($q\lambda \ll 1$), the difference in scattering amplitudes across different momentum transfers decreases algebraically as $\lambda^4$. Consequently, the experimental sensitivity to the coupling constant scales as $\alpha \propto \lambda^{-4}$ rather than exhibiting a simple exponential growth. Qualitatively, this physical boundary corresponds to the wave-mechanical resolution limit (analogous to the resolution limit in optics). In this regime, an interaction with a range significantly shorter than the neutron's de Broglie wavelength fails to induce any angularly varying phase shift and thus cannot be resolved. By employing this algebraic $\lambda^{-4}$ extrapolation, the NIST constraint naturally extends to the picometer regime where it smoothly overlaps with other exotic-atom and high-energy limits.

Consistent with Casimir force experiments, the neutron-derived constraints exhibit a power-law behavior reflecting finite-size effects (Feature~4). 
In the scattering process, the distribution of impact parameters $b$ can be interpreted as a distribution of distances $r$ between the source and probe masses. 
Although the reported constraints were derived using the Born approximation within a wave-mechanical formalization, the emergence of this common power-law feature across different experimental platforms supports this physical interpretation.

A series of experiments measuring isotope-dependent effects in atomic spectroscopy (isotope shifts) has recently been conducted to search for new bosons~\cite{PhysRevLett.134.233002, PhysRevLett.134.063002, Kyoto-Yb}. 
We have interpreted these results in terms of the coupling $\alpha$, as shown in Fig.~\ref{alpha-lambda-microscale}. 
Their current sensitivity to $\alpha$ remains relatively limited ($\alpha < 10^{29}$), primarily due to uncertainties in nuclear structure and the small mass of the electron. 
However, applying this technique to muonic or antiprotonic atoms, particularly those with simpler nuclear structures, may significantly enhance gravitational sensitivity in the future.

These studies typically report their results as the coupling strengths of a potential new boson to the electron ($y_{\rm e}$) and to the neutron ($y_{\rm n}$)~\cite{PhysRevLett.134.233002}. 
To directly compare these microscopic atomic limits with macroscopic gravity constraints, we interpret this hypothetical boson exchange as a mass-proportional force that satisfies the equivalence principle.
Under the assumption of identical coupling strengths of the boson to the neutron and proton ($y_{\rm n} = y_{\rm p}$), the coupling to a nucleus with atomic number $Z$ and neutron number $N$ (mass number $A = Z + N$) behaves coherently as $y_{\rm nuclei} = Z y_{\rm p} + N y_{\rm n} = A y_{\rm n}$.
We also neglect the nuclear mass defect and the mass difference between proton and neutron, approximating the nuclear mass as $m_{\rm nuclei} \approx A m_{\rm n}$, where $m_{\rm n}$ denotes the neutron mass.
By equating the total new boson-exchange potential (the sum of electron-proton and electron-neutron interactions) to the effective gravitational Yukawa potential, we can express $\alpha$ in terms of $y_{\rm e}$ and $y_{\rm n}$.
This yields $\alpha = y_{\rm e} y_{\rm n} / (4\pi G_{\rm N} m_{\rm n} m_{\rm e})$, where $m_{\rm e}$ denotes the electron mass.

At sub-atomic scales, nuclear or hadronic systems can be used to estimate potential gravitational contributions~\cite{0954-3899-40-3-035107}; however, the achievable precision is fundamentally limited by the theoretical uncertainties inherent in hadronic physics.
While this precision is difficult to improve significantly, the accessible length scale can be extended by increasing the energy. Indeed, high-energy collider data enable the estimation of gravitational couplings between elementary particles at scales as small as $10^{-19}$~m. 
Although the high-energy collider constraints are typically reported in the $n$--$M_{\rm D}$ parameter space, they can be converted into the $\alpha$--$\lambda$ parameters. 
We estimated the effective parameters $\delta(r)$ by fitting the model function to the converted $\Lambda(n)$ data points in Fig.~\ref{n-lambda} using the relation in Eq.~\eqref{n-l-absolute}, and subsequently plotted the corresponding $\alpha$--$\lambda$ curves in Fig.~\ref{alpha-lambda-microscale} using the exponential growth function from Eq.~\eqref{al_exp_abs}.

For the LHC constraints, we employ the systematic results from the ATLAS Collaboration~\cite{PhysRevD.103.112006}, which reported an ADD model analysis for $n = 2, \dots, 6$ (although the $n = 2$ case is currently unavailable under equivalent parameter settings from the CMS Collaboration \cite{CMS2024}).
The ATLAS results were derived in a search for large extra dimensions in monojet events, characterized by a single high-energy jet accompanied by large missing transverse momentum from undetected gravitons, in $pp$ collisions at $\sqrt{s} = 13$~TeV.

The LHC results can be compared with other experimental systems across the $\alpha$--$\lambda$ (Fig.~\ref{alpha-lambda-microscale}), $n$--$\Lambda$ (Fig.~\ref{n-lambda}), and $n$--$M_{\rm D}$ (Fig.~\ref{n-MD}) plots. In the $\alpha$--$\lambda$ parametrization used to search for new boson exchange forces, the LHC provides the most stringent constraints for $\lambda < 10$~pm, a regime where atomic and nuclear tests are attempted. 
Regarding the large extra-dimension model, Fig.~\ref{n-lambda} allows for a direct comparison of the results. For most values of $n$, the LHC establishes the strongest constraints on $\Lambda$.

For the specific case of $n = 2$, the constraints from high-energy collider experiments and short-range tabletop experiments are highly competitive when compared using the consistent phenomenological scale parameter $\Lambda$ defined in Eq.~\eqref{V_nl}. 
The HUST tabletop experiment provides a lower limit of $\Lambda < 11\,\upmu\text{m}$~\cite{PhysRevLett.124.051301}, which represents the characteristic transition scale where the inverse-square law begins to be violated. 
Using the direct conversion relation in Eq.~\eqref{lambda-MD}, the LHC 2021 limit of $M_{\rm D} > 11.2$~TeV~\cite{PhysRevD.103.112006} yields an equivalent scale limit of $\Lambda < 4\,\upmu\text{m}$, which is plotted directly in Fig.~\ref{n-lambda}. 
In our previous review~\cite{Murata_2015}, the UW constraint stood at $\Lambda < 23\,\upmu\text{m}$, while the LHC limit was $\Lambda < 130\,\upmu\text{m}$. 
It is worth noting that while these parameters are plotted consistently as the phenomenological transition scale $\Lambda$, a geometric interpretation of $\Lambda$ as the physical size of the extra dimension depends on the choice of compactification conventions. If $\Lambda$ is interpreted as the torus circumference ($\Lambda \equiv 2\pi R$), the physical compactification radius corresponds to $R \approx 0.64\,\upmu\text{m}$ for the LHC and $R \approx 1.75\,\upmu\text{m}$ for HUST.

A similar comparison is presented in the $n$--$M_{\rm D}$ plane (Fig.~\ref{n-MD}), where the limits are plotted consistently using the algebraic relation in Eq.~\eqref{ADD} directly from the transition scale parameter $\Lambda$. 
Under this unified phenomenological parameter space, HUST and the LHC provide lower limits of $M_{\rm D} > 6.6$~TeV and $M_{\rm D} > 11.2$~TeV for $n = 2$, respectively. 
By contrast, in our previous review~\cite{Murata_2015}, the UW tabletop limit was reported as $M_{\rm D} > 4.6$~TeV, while the LHC limit was $M_{\rm D} > 1.9$~TeV. 
Nevertheless, under any consistent parameter definition, both experimental frontiers remain highly competitive and complementary to each other in the search for extra dimensions.

\backmatter


\section{Conclusion}\label{conclusion}
We have performed a comprehensive and unified analysis comparing short-range gravity constraints across the $\alpha$--$\lambda$, $n$--$\Lambda$, and $n$--$M_{\rm D}$ parameter spaces. 
This consistent framework successfully demonstrates how diverse experimental data—ranging from high-energy collider searches to tabletop precision force measurements—can be mapped and compared within the context of large extra-dimension models. 
Our analysis shows that in the unified phenomenological parameter space for $n = 2$, the latest LHC results set a constraint on the transition scale of $\Lambda < 4\,\upmu\text{m}$, while the micrometer-scale torsion-balance experiment by the HUST group establishes an exceptionally competitive limit of $\Lambda < 11\,\upmu\text{m}$.

Importantly, laboratory tests of the gravitational inverse-square law remain uniquely competitive and complementary to high-energy collider constraints at around the micrometer scale for the $n = 2$ case. 
Looking forward, the development of novel experimental techniques, alongside significant advancements in the theoretical calculation of background electromagnetic couplings, will pave the way for high-precision gravity tests at sub-micrometer scales.

\bmhead{Acknowledgements}
The authors are grateful to S. Mukohyama for encouraging this research and for providing insightful theoretical discussions. 
We also thank Y. Kamiya for valuable discussions regarding the interpretation of neutron scattering data.
Finally, we thank the members of the Rikkyo group---K. Ninomiya, S. Zeidler, N. Kobayashi, M. Yokomizo, Y. Ishikawa, S. Akamatsu, T. Ito, and H. Sato---for their contributions to the discussions on gravity physics. 

\bmhead{Author Contributions}
T. Fujiie and S. Suzuki contributed to the literature search and the interpretation of the experimental data. 
J. Murata supervised the research, performed the data analysis, synthesized the results, and drafted the manuscript as the corresponding author.

\bmhead{Declarations}

\bmhead{Funding}
This work was supported by JSPS KAKENHI Grant-in-Aid for Scientific Research (B), Grant Numbers JP23K20851 and JP26K00713. 

\bmhead{Availability of data and materials}
The source graphical files for the constraint curves presented in this study are available from the corresponding author upon reasonable request.

\bmhead{Competing interests}
The authors declare that they have no competing interests.


\bibliography{gravity}

\end{document}